\documentclass[epj]{svjour}
\usepackage{graphics}
\usepackage{graphicx}
\usepackage{wrapfig}
\usepackage{amssymb}
\usepackage{amsfonts}
\usepackage{amsmath}
\usepackage{longtable}
\usepackage{rotating}
\usepackage{lscape}
\usepackage{epsfig}
\usepackage{multirow}
\usepackage{textcomp}
\usepackage{gensymb}
\usepackage{multirow}
\usepackage{array}
\usepackage{url}
\usepackage[T1,T2A]{fontenc}
\usepackage{float}
\usepackage{xcolor}
\usepackage{enumitem}
\usepackage{makecell}
\usepackage{mwe,tikz}
\usepackage{wrapfig}
\usepackage[percent]{overpic}
\usepackage{hyperref}
\hypersetup{
  colorlinks  = true,
  urlcolor = blue,
  linkcolor = blue,
  citecolor = blue
}
\usepackage{orcidlink}

\usepackage[sorting=none, backend=biber, style=numeric]{biblatex}
\begin{document}

\title{Spontaneous fission of $^{246}$Fm}

\author{A.V. Isaev\inst{1}\thanks{Corresponding author: isaev@jinr.ru}\orcidlink{0000-0002-6064-4065} \and 
R.S. Mukhin\inst{1}\orcidlink{0000-0002-1291-995X} \and
A.V. Andreev\inst{1} \and
M.A. Bychkov\inst{1} \and
M.L. Chelnokov\inst{1}\orcidlink{0000-0003-0300-6371} \and
V.I. Chepigin\inst{1} \and 
H.M. Devaraja\inst{1}\orcidlink{0000-0002-6941-4143} \and
O. Dorvaux\inst{2}\orcidlink{0000-0002-5838-9042} \and
M. Forge\inst{2}\orcidlink{0000-0003-3185-2313} \and
B. Gall\inst{2}\orcidlink{0000-0002-9858-1397} \and
K. Hauschild\inst{3}\orcidlink{0000-0003-2862-2445} \and 
I.N. Izosimov\inst{1}\orcidlink{0000-0003-2744-3453} \and 
K. Kessaci\inst{2}\orcidlink{0000-0002-3608-3587} \and 
A.A. Kuznetsova\inst{1} \and  
A. Lopez-Martens\inst{3}\orcidlink{0000-0001-9985-6228} \and 
O.N. Malyshev\inst{1,4} \and 
A.G. Popeko\inst{1,4} \and 
Yu.A. Popov\inst{1,4}\orcidlink{0000-0001-6190-3340} \and 
A. Rahmatinejad\inst{1}\orcidlink{0000-0001-5310-133X} \and 
B. Sailaubekov\inst{1,5,6}\orcidlink{0000-0001-7915-6021} \and 
T.M. Shneidman\inst{1,7}\orcidlink{0000-0003-3226-8768} \and 
E.A. Sokol\inst{1} \and 
A.I. Svirikhin\inst{1,4}\orcidlink{0000-0002-4033-7076} \and 
D.A. Testov\inst{1}\orcidlink{0000-0003-3193-5311} \and 
M.S. Tezekbayeva\inst{1,5}\orcidlink{0000-0002-0508-2274} \and 
A.V. Yeremin\inst{1,4}\orcidlink{0000-0001-5092-8489} \and 
N.I. Zamyatin\inst{1} \and 
K.Sh. Zhumadilov\inst{6}\orcidlink{0000-0002-0205-4585}}

\institute{$^{1}$\,Joint Institute for Nuclear Research, Dubna, Russia\\
$^{2}$\,Université de Strasbourg, CNRS, Strasbourg, France\\
$^{3}$\,IJCLab, IN2P3-CNRS, Université Paris-Saclay, Orsay, France\\
$^{4}$\,Dubna State University, Dubna, Russia\\
$^{5}$\,Institute of Nuclear Physics, Almaty, Kazakhstan\\
$^{6}$\,L.N. Gumilyov Eurasian National University, Nur-Sultan, Kazakhstan\\
$^{7}$\,Kazan Federal University, Kazan, Russia}

\abstract{
An experiment on the study of the $^{246}$Fm spontaneous fission was conducted using the SHELS separator. The isotope was synthesized in the complete fusion reaction of $^{40}$Ar beam ions and $^{208}$Pb target nuclei.
The neutron yields of $^{246}$Fm spontaneous fission ($\overline{\nu} = 3.79\pm0.30$, $\sigma^{2}_{\nu} = 2.1$) were obtained using the SFiNx detector system. The multiplicity distribution of emitted prompt neutrons was restored using the Tikhonov method of statistical regularisation ($\overline{\nu}_{r} = 3.79\pm0.20$, $\sigma^{2}_{\nu r} = 2.8$). The spontaneous fission branching ratio ($b_{SF} = 0.061\pm0.005$) and the half-life ($T_{1/2} = 1.50^{+0.08}_{-0.07}$ s) of the isotope were determined.
The experimental data were compared with scission point model predictions. Excellent convergence was observed in the average number of neutrons per spontaneous fission process. However, the forms of the experimental and model prompt neutron multiplicity distributions differ significantly.
} 

\maketitle

\label{sec:introduction}
\section{Introduction}

Spontaneous fission is a common decay mode for heavy atomic nuclei and defines the limits of existence of chemical elements in this region. The mass and energy distributions of fission fragments and prompt neutron yields are important characteristics describing this process. However, the study of the properties of the heaviest nuclei is complicated by their short lifetimes and small formation cross-sections, which requires the use of advanced experimental methods.

Data on prompt neutron yields from spontaneous fission were accumulated earlier for many isotopes with $Z < 100$, mainly in offline experiments. For fermium ($Z = 100$), such data were obtained both for neutron-rich isotopes $^{254,256,257}$Fm \cite{GindlerFm254, SokolFm256, HoffmanFm257} and neutron-deficient ones (online experiments) $^{244,246}$Fm \cite{SvirikhinFm244, SvirikhinFm246}. The $^{246}$Fm prompt $\gamma$-ray spectroscopy was performed at the University of Jyv{\"a}sjyl{\"a} \cite{Piot2012} and is the lowest cross-section reaction where this was possible for such heavy nuclei. The $^{246}$Fm prompt neutron yields are the most poorly known because only 108 \cite{SvirikhinFm246} fission events were obtained. Therefore, the purpose of the present work was to repeat the synthesis of $^{246}$Fm in order to obtain more statistics and refine its prompt neutron yield data.

\label{sec:experiment}
\section{Experiment}

\begin{figure}
\resizebox{0.5\textwidth}{!}{
  \includegraphics{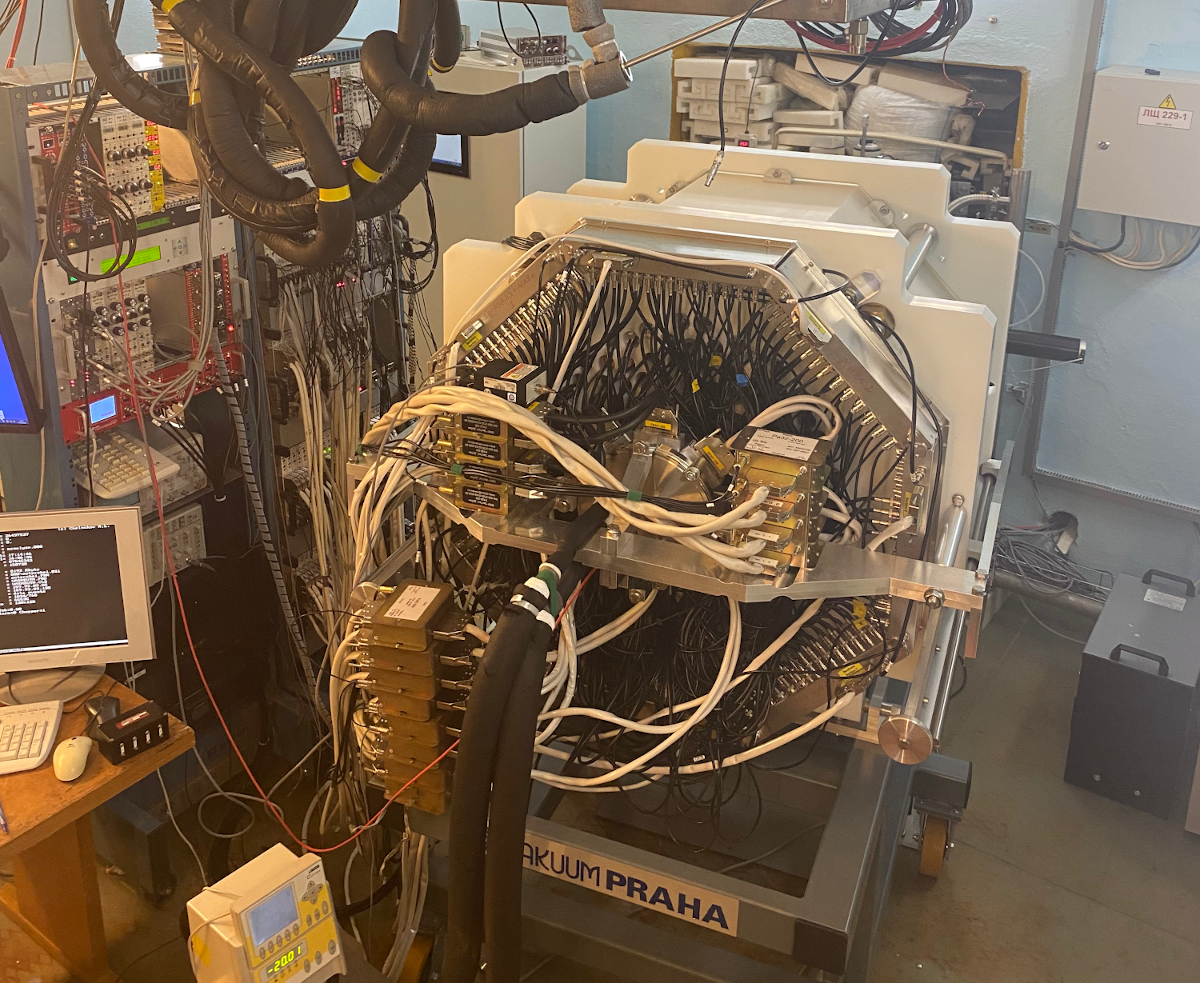}
}
\caption{Photograph of the SFiNx detector system at the focal plane of the SHELS.}
\label{sfinx}
\end{figure}

\begin{figure}
\resizebox{0.5\textwidth}{!}{ 
  \includegraphics{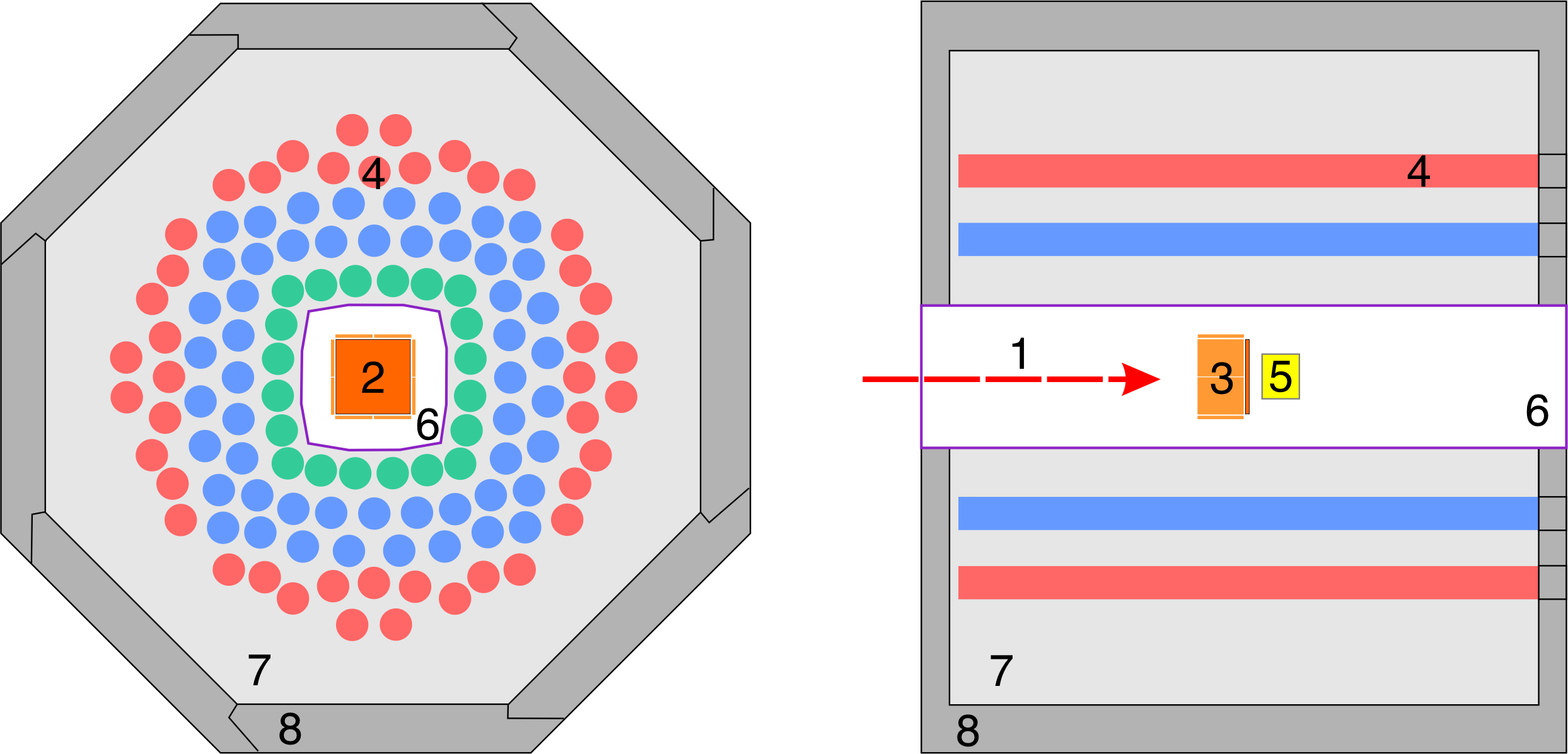}
}
\caption{The SFiNx scheme (front view -- left; side view -- right). The legend: 1 -- recoil nuclei; 2 -- focal plane Si-detector; 3 -- tunnel Si-detector array; 4 -- $^{3}$He-counters; 5 -- scintillator detector; 6 -- vacuum chamber; 7 -- moderator; 8 - shield.}
\label{scheme}
\end{figure}

An experiment aimed at studying the spontaneous fission properties of the $^{246}$Fm isotope was carried out at FLNR JINR using the SHELS separator \cite{PopekoSHELS} and the SFiNx detection system \cite{SFiNx} (Fig.~\ref{sfinx}). The system includes 116 $^{3}$He-neutron counters that allow registering multiple prompt neutrons emitted in the spontaneous fission process of the nucleus, as well as an assembly (<<well>> like) of double-sided Si detectors with a 128$\times$128-strip focal-plane detector and 8 tunnel 16$\times$16-strip detectors for fission-fragment and $\alpha$-particle registration (Fig.~\ref{scheme}). The high granularity of the SFiNx neutron detector makes it possible to register multiple neutron events with a negligible probability that a single $^{3}$He-counter will detect several neutrons simultaneously within the coincidence time window \cite{SFiNx}. The neutron registration efficiency, measured with a $^{248}$Cm source, is (54.8$\pm$0.1)\%, and the average neutron lifetime in the array is 18.4$\pm$0.2 $\mu$s. The detection efficiency of the focal-plane Si detector for $\alpha$-particles emitted by implanted nuclei is $\sim$50\% and 100\% for at least one of the two fission fragments.

The complete fusion reaction of $^{40}$Ar$^{16+}$ ions from the U-400 cyclotron with a $^{208}$Pb target nuclei was used to synthesize $^{246}$Fm. The PbS target thickness was 450 $\mu$g/cm$^{2}$ (the $^{208}$Pb isotope enrichment was over 99\%), and a 1.5 $\mu$m thick titanium backing was used. The ion beam energy was \mbox{183$\pm$3 MeV} and it was chosen close to the maximum of the excitation function of the $2n$-evaporation channel \cite{SvirikhinFm246, Oganessian1975, Piot2012}. The total number of beam ions that passed through the target and then stopped in a faraday cup was about 1.7$\times$10$^{18}$.

Since $^{206}$Pb impurities in the target (< 1\%) can lead to the production of the spontaneously fissile $^{244}$Fm with T$_{1/2}\sim$ 3 ms, fission fragments from $^{246}$Fm were searched for in the time interval 30 ms -- 15.4 s following the registration of the implanted reaction products.

\begin{figure}
\resizebox{0.5\textwidth}{!}{
  \includegraphics{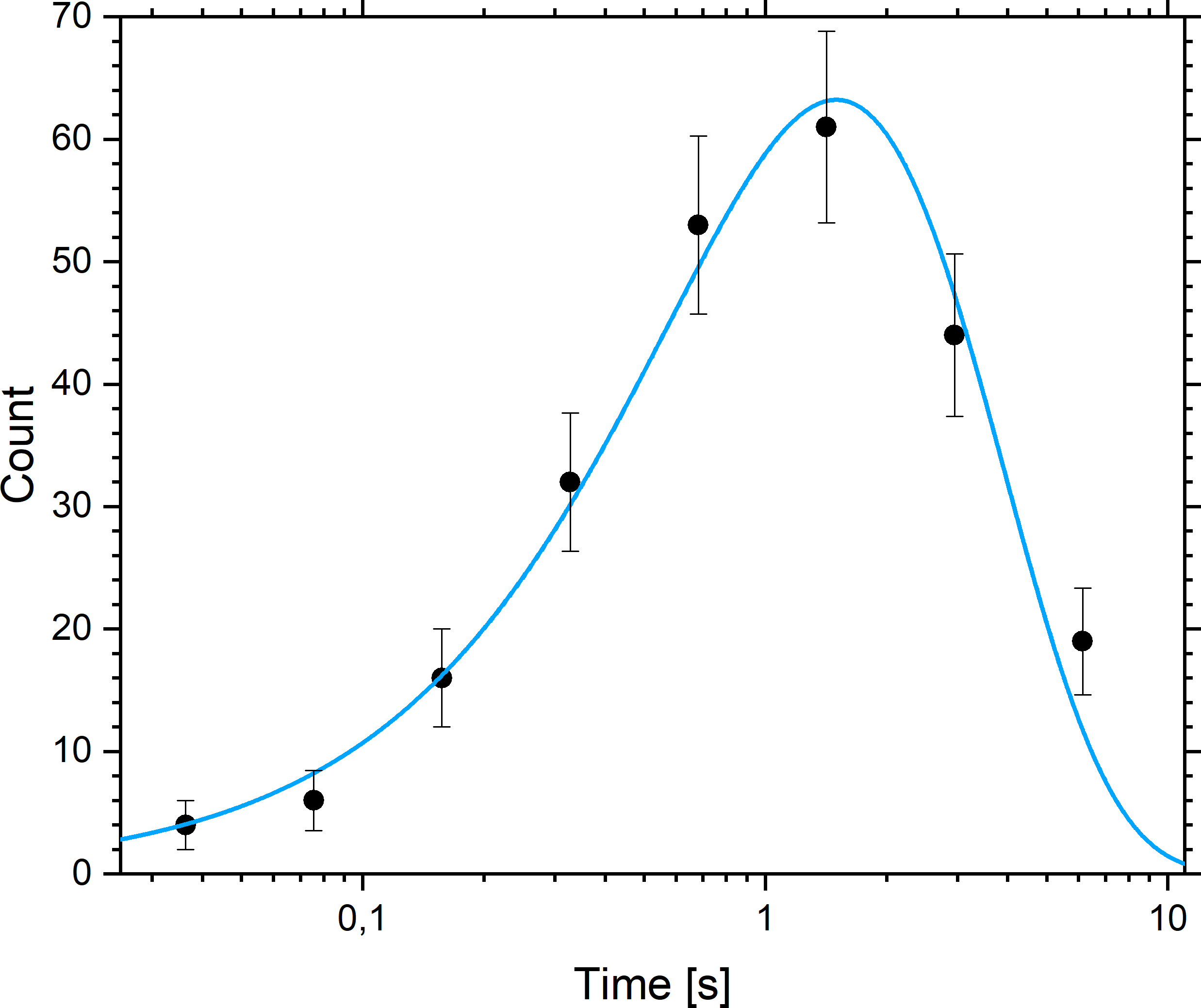}
}
\caption{Distribution of time differences between recoil nuclei and fission fragments registration for $^{246}$Fm (dots) and fitting by method \cite{Schmidt2000} (curve).}
\label{fissiontimes}
\end{figure}

A total of 235 $^{246}$Fm spontaneous fissions were found during data analysis, and only 2 events (in the interval 0 -- 30 ms) were assigned to the spontaneous fission of $^{244}$Fm. The $^{246}$Fm half-life was obtained as $T_{1/2} = 1.50^{+0.08}_{-0.07}$ s from the time distribution shown in Fig.~\ref{fissiontimes}. The value is in good agreement with the previously-measured values of 1.54$\pm$0.04 s \cite{Venhart2011}, 1.3$\pm$0.2 s \cite{SvirikhinFm246} and 1.6$\pm$0.2 s \cite{Piot2012}.

\begin{figure}
\resizebox{0.5\textwidth}{!}{
   \includegraphics{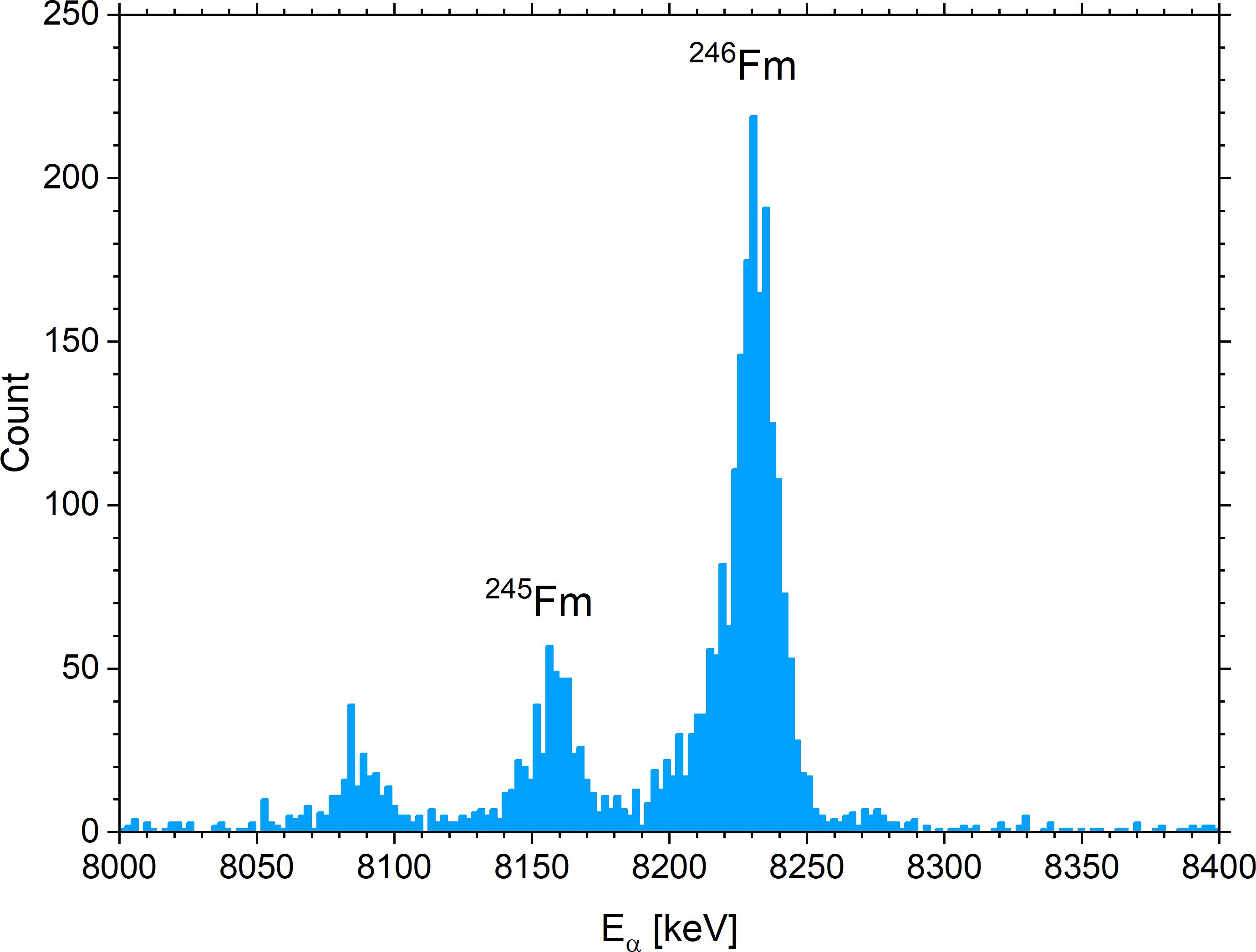} 
}
\caption{Spectrum of $\alpha$-particles obtained in the search for correlations with implanted recoil nuclei.}
\label{alpha}
\end{figure}

To determine the spontaneous fission branching ratio, $\alpha$ particles were searched for in the time interval 30 ms -- 15.4 s from the recoil implantation signal (Fig.~\ref{alpha}). After background subtraction, 1809 $\alpha$-decays were found in the energy range 8190 -- 8260 keV. The spontaneous fission branching ratio was obtained as $b_{SF} = 0.061\pm0.005$, which agrees with the previously published values of 0.068$\pm$0.006 \cite{Venhart2011} and 0.05$\pm$0.03 \cite{SvirikhinFm246}.

\begin{figure}
\resizebox{0.5\textwidth}{!}{
   \includegraphics{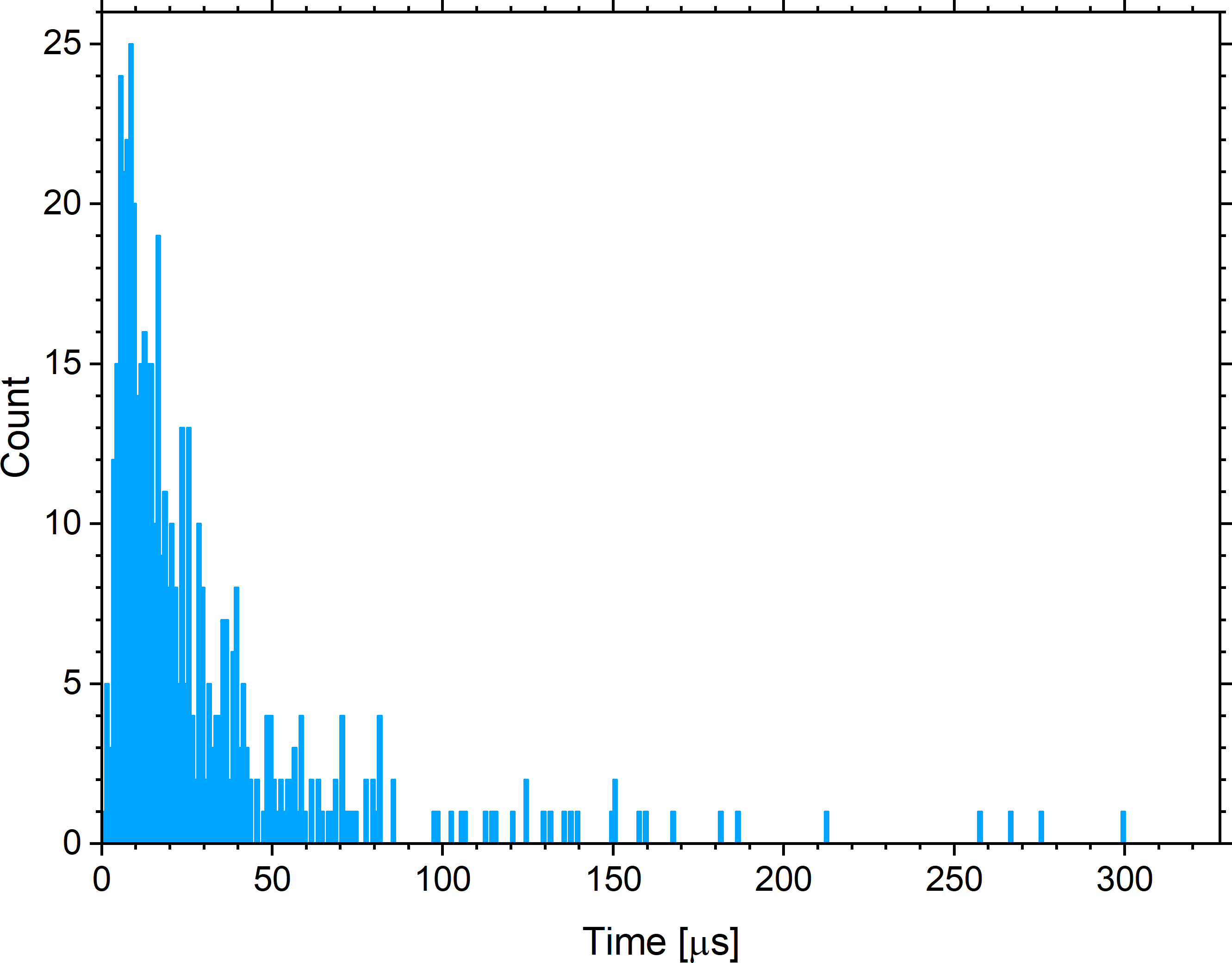}
}
\caption{Distribution of time differences between neutron detection and the spontaneous fission of $^{246}$Fm.}
\label{neutrontimes}
\end{figure}

Prompt neutrons emitted in the spontaneous fission of fermium were searched for in the time interval 0 -- 128 $\mu$s from the moment of the fission-fragment registration. The neutron time distribution in the coincidence window is shown in Fig.~\ref{neutrontimes}. A total of 488 prompt neutrons from 235 $^{246}$Fm spontaneous fission events were registered.

\begin{figure}
\resizebox{0.5\textwidth}{!}{
   \includegraphics{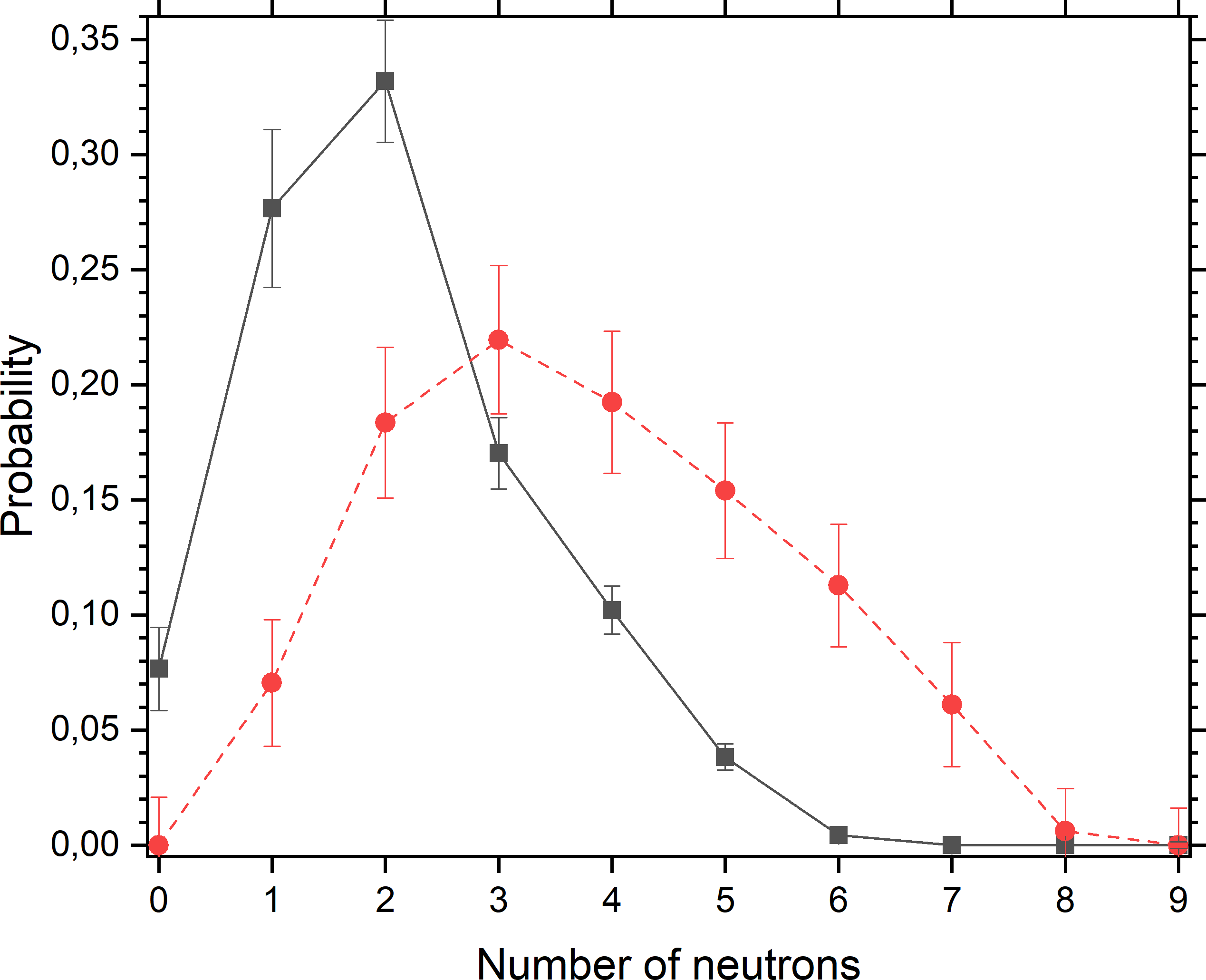}
}
\caption{Multiplicity distributions of the neutrons emitted in the spontaneous fission of $^{246}$Fm: detected in the experiment (squares) and reconstructed (circles). The lines connecting points have been added for clarity.}
\label{multiplicity}
\end{figure}

The prompt neutron multiplicity for $^{246}$Fm extracted from the data is shown in Fig.~\ref{multiplicity}. Taking into account the detector efficiency, the characteristics of the multiplicity distribution of the emitted neutrons were obtained: the mean as $\overline{\nu} = 3.79\pm0.30$ and the variance as $\sigma^{2}_{\nu} = 2.1$. The obtained $\overline{\nu}$ value agrees well with 3.55$\pm$0.50 already measured in Ref. \cite{SvirikhinFm246}.

The true form of the neutron multiplicity distribution was restored using Tikhonov's method of statistical regularization \cite{MukhinRegularisation}. The reconstruction results are shown in Fig.~\ref{multiplicity} and in the table~\ref{tab:fm246_multiplicity}.

For the reconstructed distribution, the average number of neutrons is 3.79$\pm$0.20 and the variance is 2.8.

An additional search was made for neutron events in the time interval 200 -- 328 $\mu$s from the $^{246}$Fm spontaneous fission registration for background estimation. The interval duration was equal to the duration of the $^{246}$Fm prompt neutrons search. The choice of the lower boundary of the search interval was based on the average neutron lifetime in the detector and set to eliminate the detection of fermium prompt neutrons. Only 5 background neutrons were detected (see Fig.~\ref{neutrontimes}) for 235 $^{246}$Fm spontaneous fission events. The influence of the background was insignificant in comparison with the level of statistical errors obtained in the experiment (the background to signal ratio is about 1\% whereas the relative statistical error is about 8\%).

\begin{table}
\caption{Prompt neutron emission probability for $^{246}$Fm. The legend: $n$ -- number of neutrons; $P_{m}$ -- measured value; $\Delta P_{m}$ -- error of measured value; $P_{r}$ -- reconstructed value; $\Delta P_{r}$ -- error of reconstructed value.}
\renewcommand{\arraystretch}{1.5}
\centering
\begin{tabular}{c|c|c|c|c}
\hline
\thead{$n$} & \thead{$P_{m}$} & \thead{$\Delta P_{m}$} & \thead{$P_{r}$} & \thead{$\Delta P_{r}$} \\ \hline
0 & 0.077 & 0.018 & 0 & 0.021 \\ \hline
1 & 0.277 & 0.034 & 0.070 & 0.028 \\ \hline
2 & 0.332 & 0.027 & 0.184 & 0.033 \\ \hline
3 & 0.170 & 0.016 & 0.220 & 0.032 \\ \hline
4 & 0.102 & 0.010 & 0.192 & 0.031 \\ \hline
5 & 0.038 & 0.006 & 0.154 & 0.030 \\ \hline
6 & 0.004 & 0.002 & 0.113 & 0.027 \\ \hline
7 & 0 & $<$ 0.002 & 0.061 & 0.027 \\ \hline
8 & -- & -- & 0.006 & 0.018 \\ \hline
9 & -- & -- & 0 & 0.016 \\ \hline
\end{tabular}
\label{tab:fm246_multiplicity}
\end{table}

\label{sec:model}
\section{Model}

The theoretical calculations of neutron multiplicity distributions were carried out with the improved scission point model \cite{Andreev2006}. The model assumes that the observed characteristics of the fission process are formed at the scission point, where the fissile nucleus can be represented as a superposition of systems consisting of two adjoining fragments $(A_1,Z_1,\beta_1)$ and $(A_2,Z_2,\beta_2)$. Assuming statistical equilibrium, the probability of realization of various binary configurations $G(A_i,Z_i,\beta_i,i=1,2)$ is determined by the potential energy as a function of masses $A_i$, charges $Z_i$, deformations $\beta_i$ of the fragments \cite{Adamian1996} and excitation energy $U^\star$. For a particular scission configuration with fixed charge and mass numbers of the fragments, the potential energy as a function of the deformation parameters $\beta_1$ and $\beta_2$ can have one or several minima depending on the interplay of the macroscopic liquid-drop energy and microscopic shell effects. Calculations show that for the nuclei considered in this paper, the potential energy minimum occurs at deformations $\beta_i \gtrsim 2$. For such large deformations, the pocket in the interaction potential of the fragments disappears, and the system becomes unstable. Thus, the fissile nucleus reaches strongly deformed configurations with fewer probabilities than it follows from the assumption of statistical equilibrium. To account for this effect, the probabilities $G(A_i,Z_i,\beta_i)$ are multiplied by a factor
\begin{eqnarray}
\prod_{i=1,2}\frac{1}{1+\exp{\frac{\beta_i-\beta_0}{s}}}.
\label{correction}
\end{eqnarray}
In this work, for all nuclei, the parameter values $\beta_0=1.7$ and $s=0.08$ were used.

For each system with given masses, charges, and deformations of the fragments, the probability of $n$ neutrons being emitted is calculated as
\begin{eqnarray}
\begin{split}
P(n)=\sum_{x=1}^{n}\int_{0}^{U^\star}F(U_{1}) P_{1}(x, U_{1}+U_{1}^{d})\\P_{2}(n-x, U_{2}+U_{2}^{d}) dU_{1}, 
\end{split}
\end{eqnarray}
where $U_i$ and $U_i^d$, ($i=1,2$) are the excitation energy of the $i$-th fragment available at the scission point and its deformation energy with respect to its ground state, respectively. The quantity $P_{i}(x, U_{i}+U_{i}^{d})$ determines the probability that exactly $x$ neutrons will be emitted from the $i$-th fragment \cite{Vandenbosch1973}. The probability that the excitation energy at the scission point will be distributed between the fragments as $U_1$ and $U_2$, $U_{1}+U_{2}=U^{\star}$, is given by the function
\begin{eqnarray}
F(U_{1}) \sim \rho(a_{1},U_{1})\rho(a_{2},U_2=U^\star-U_{1}),
\end{eqnarray}
where the level densities $\rho$ of the fragments are taken in the form of a Fermi gas distribution with the level density parameters $a_i=A_i/12$ \cite{Ignatyuk1983}. After separation, it is assumed that the deformation energy of the fragment is converted into its excitation energy.

\begin{figure}
\resizebox{0.5\textwidth}{!}{
   \includegraphics{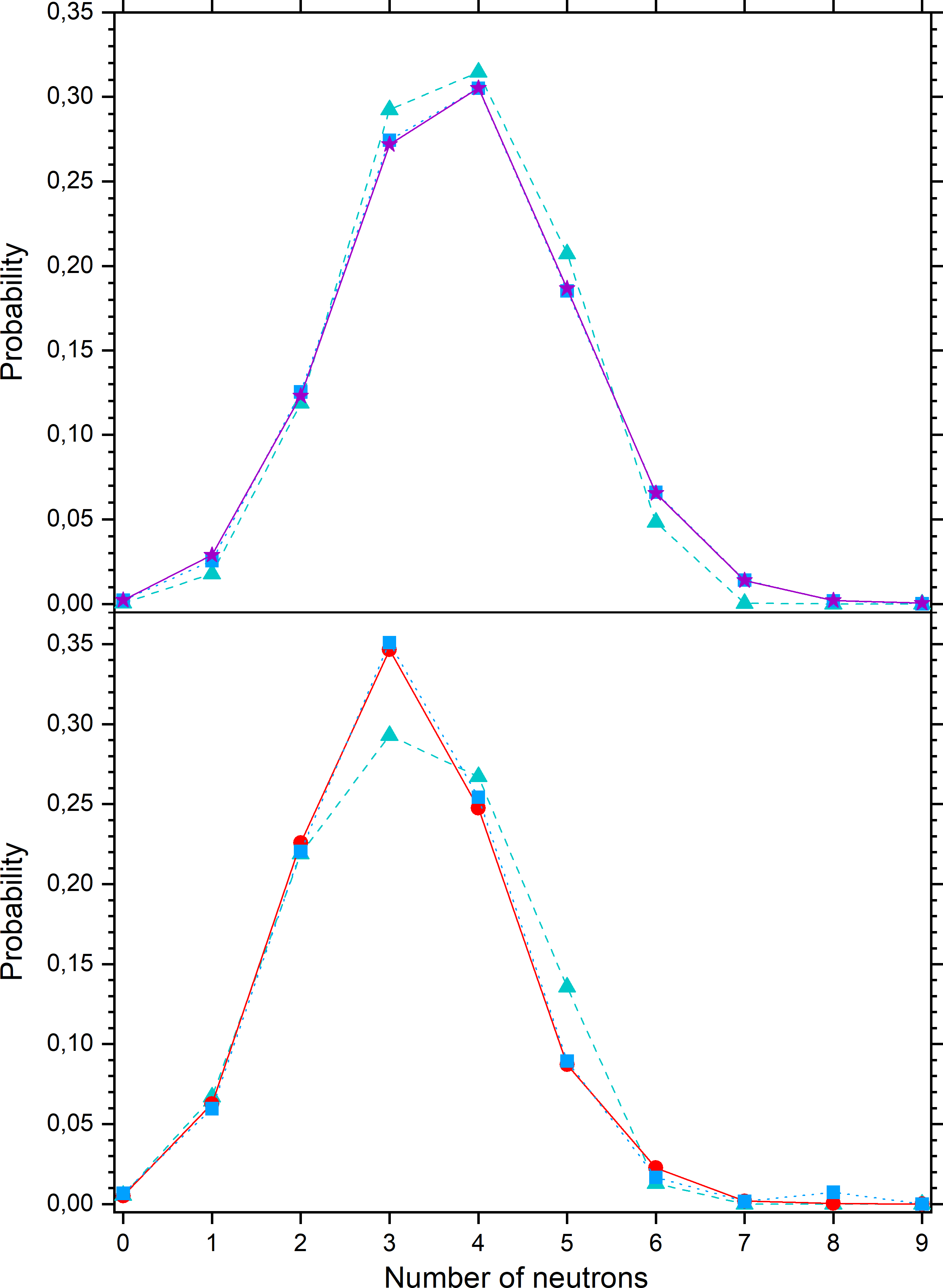}
}
\caption{Prompt neutron multiplicity distributions for $^{252}$Cf (top) and $^{248}$Cm (bottom). Symbols: triangles -- calculations within the scission point model; squares -- data from \cite{Holden1986}; stars -- data from \cite{Vorobyev2005}; circles -- current work.}
\label{Cf_and_Cm}
\end{figure}

Within the framework of this model, the distributions of prompt neutron multiplicities for $^{252}$Cf and $^{248}$Cm were calculated (Fig.~\ref{Cf_and_Cm}). One can observe a reasonable agreement between the model predictions and experimental data for both the average neutron numbers and the shape of the distributions (table~\ref{tab:model_vs_experiments}).

\begin{figure}
\resizebox{0.5\textwidth}{!}{
    \includegraphics{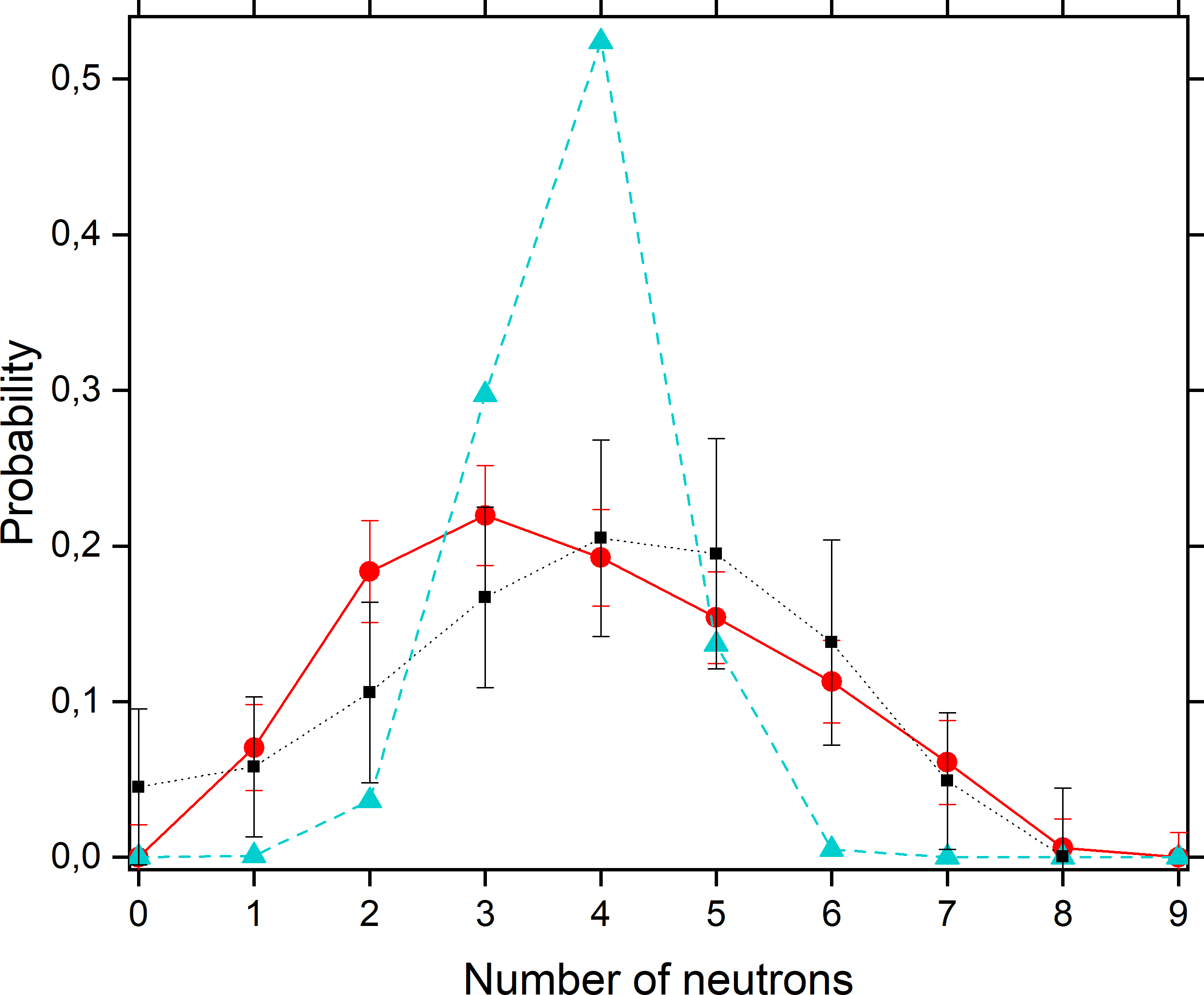}
}
\caption{Reconstructed prompt neutron multiplicity distributions for $^{246}$Fm (circles -- values obtained in this work; squares -- values taken from \cite{MukhinRegularisation}) and calculation made within the scission point model (triangles). The lines have been added to guide the eye.}
\label{model}
\end{figure}

In Fig.~\ref{model}, the calculation results are presented for $^{246}$Fm. The calculated average number of neutrons in the spontaneous fission process is 3.77, which agrees very well with the value measured in this experiment 3.79$\pm$0.30. However, the calculated distribution for $^{246}$Fm is narrower than the experimental one (the numerical parameters of the distributions are given in table~\ref{tab:model_vs_experiments}). The discrepancies between the theoretical and experimental distributions exceed the values of statistical errors.

\begin{figure}
\resizebox{0.5\textwidth}{!}{
   \includegraphics{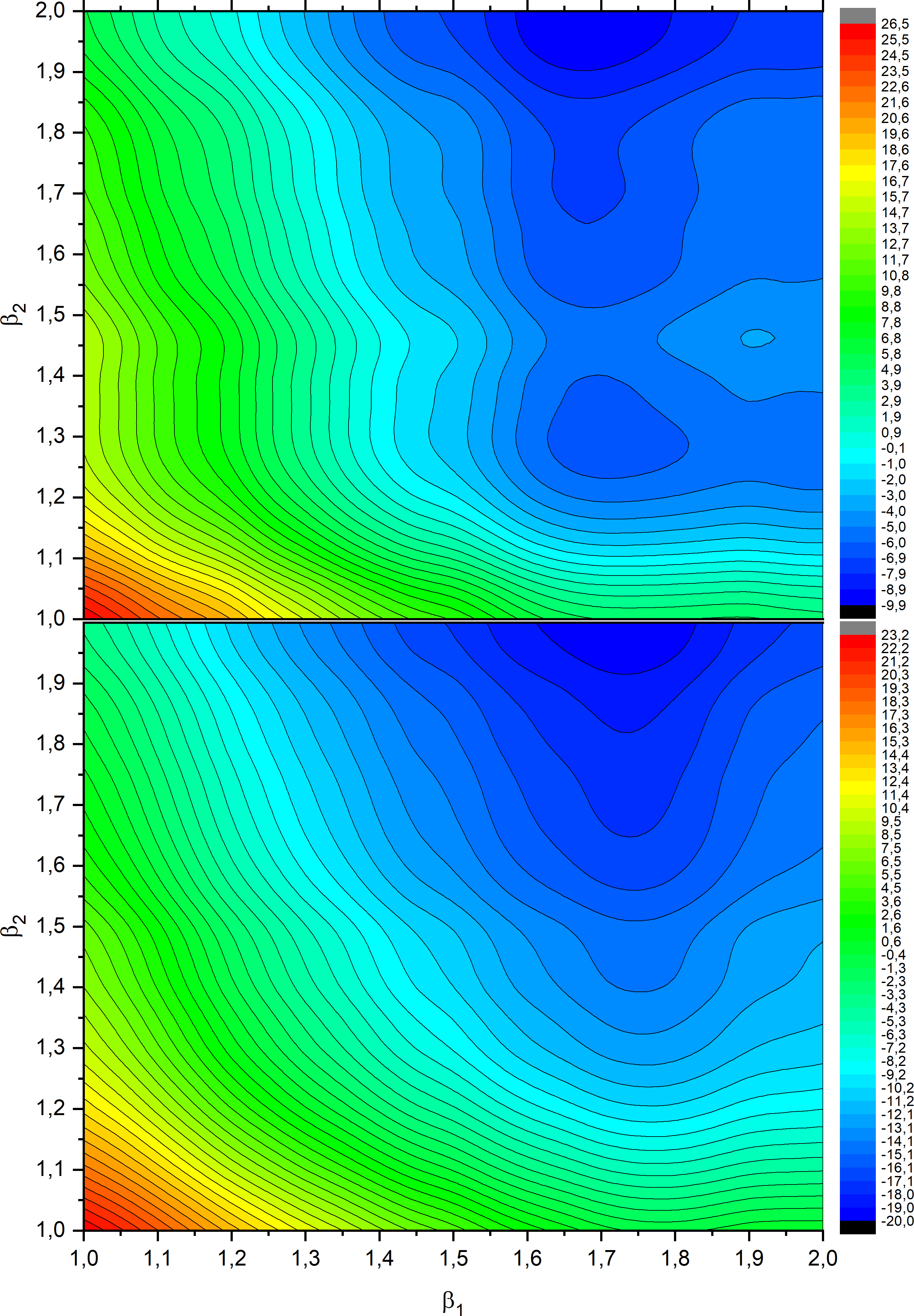}
}
\caption{Potential energy of scission configurations $^{108}$Mo$+$$^{140}$Xe (top) and $^{102}$Mo$+$$^{144}$Ce (bottom) as a function of the fragment deformations. These configurations have the maximal yields in the fission of $^{248}$Cm and $^{246}$Fm, respectively.}
\label{Cm_and_Fm_surfaces}
\end{figure}

A comparison of the theoretical distributions for $^{252}$Cf, $^{248}$Cm, and $^{246}$Fm show that for the latter there is a strong suppression of low multiplicity events. To understand this, a comparative analysis of the corresponding potential energy surfaces has been conducted (Fig.~\ref{Cm_and_Fm_surfaces}). We observed that for the most probable scission configurations of $^{252}$Cf and $^{248}$Cm, the potential energy as a function of the fragment's deformation has two minima; a more compact one ($\beta_1 \approx 1.7, \beta_2 \approx 1.3$) and a strongly deformed one ($\beta_1 \approx 1.7, \beta_2 \gtrsim 2$). The compact minimum leads to larger total kinetic energy of the fragments and a smaller number of emitted neutrons. In the statistical equilibrium assumption, the probability of realization of different deformations is split into two sizeable parts corresponding to these two minima. The use of the factor from equation \eqref{correction} which accounts for deviations from statistical equilibrium increases the role of the compact minimum.

However, in the case of $^{246}$Fm, the compact minimum in the potential energy as a function of deformation is absent. The probability distribution taken in the statistical equilibrium approach is fully concentrated in the region of large deformation. The compact systems related to a small number of emitted neutrons are realized with negligible probabilities. This cannot be significantly corrected by the factor from equation \eqref{correction}. Therefore, we can conclude that a more accurate account for the non-equilibrium effects is needed for $^{246}$Fm. This work is currently in progress.

\begin{table}
\caption{Comparison of experimental and model prompt neutron multiplicities distributions  for $^{248}$Cm, $^{252}$Cf and $^{246}$Fm}
\renewcommand{\arraystretch}{1.5}
\centering
\begin{tabular}{c|c|cclc|c}
\hline
\multirow{2}{*}{Isotope}     & \multirow{2}{*}{Property} & \multicolumn{4}{c|}{Experiments} & \multirow{2}{*}{\begin{tabular}[c]{@{}c@{}}Model \\ prediction\end{tabular}} \\ \cline{3-6}
                            &                                 & \multicolumn{1}{l|}{\cite{Holden1986}}    & \multicolumn{1}{l|}{\cite{Vorobyev2005}}    & \multicolumn{1}{l|}{\cite{MukhinRegularisation}}    & \multicolumn{1}{l|}{This work} &                                                                                    \\ \hline
\multirow{2}{*}{$^{248}$Cm} & $\overline{\nu}$                & \multicolumn{1}{c|}{3.13} & \multicolumn{1}{c|}{3.13} & \multicolumn{1}{l|}{--}   & 3.13                               & 3.21                                                                               \\ 
                            & $\sigma^{2}_{\nu}$              & \multicolumn{1}{c|}{1.29} & \multicolumn{1}{c|}{1.37} & \multicolumn{1}{l|}{--}   & 1.35                               & 1.42                                                                               \\ \hline
\multirow{2}{*}{$^{252}$Cf} & $\overline{\nu}$                & \multicolumn{1}{c|}{3.76} & \multicolumn{1}{c|}{3.76} & \multicolumn{1}{l|}{--}   & --                                 & 3.72                                                                               \\ 
                            & $\sigma^{2}_{\nu}$              & \multicolumn{1}{c|}{1.59} & \multicolumn{1}{c|}{1.62} & \multicolumn{1}{l|}{--}   & --                                 & 1.26                                                                               \\ \hline
\multirow{2}{*}{$^{246}$Fm} & $\overline{\nu}$                & \multicolumn{1}{c|}{--}   & \multicolumn{1}{c|}{--}   & \multicolumn{1}{l|}{3.9} & 3.79                               & 3.77                                                                               \\
                            & $\sigma^{2}_{\nu}$              & \multicolumn{1}{c|}{--}   & \multicolumn{1}{c|}{--}   & \multicolumn{1}{l|}{3.1} & 2.80                               & 0.57                                                                               \\ \hline
\end{tabular}
\label{tab:model_vs_experiments}
\end{table}

\label{sec:conclusions}
\section{Сonclusions}

As a result of the experiment, the emission probabilities of $^{246}$Fm prompt neutrons of different multiplicities were measured with the best accuracy (Fig.~\ref{model}). The most accurate value of the average number of neutrons per spontaneous fission decay of $^{246}$Fm is obtained $\overline{\nu} = 3.79\pm0.30$.

Since the neutron detection efficiency of the SFiNx \cite{SFiNx} is (54.8$\pm$0.1)\% and the collected statistics is small (235 spontaneous fission events with 488 prompt neutrons in total), in order to obtain the true form of the prompt neutron multiplicity distribution, the use of the statistical regularization method is necessary. This procedure was successfully made for $^{246}$Fm prompt neutrons data and as a result, the true shape of the emitted neutron multiplicity distribution was obtained with the best precision.

Today there is no complete model of nuclear fission that could describe all the characteristics of the fission process for each specific nucleus. For the development of theoretical approaches, it is important to obtain new experimental data, especially for short-lived nuclei in the $Z \geq 100$ region. In this paper, we have used an improved scission point model \cite{Andreev2006}, which made it possible to predict prompt neutron multiplicity distributions for $^{252}$Cf, $^{248}$Cm, and $^{246}$Fm nuclei. The model perfectly describes the average number of neutrons in the fission processes of all the listed nuclei as well as the shapes of the neutron multiplicity distributions for $^{252}$Cf and $^{248}$Cm. For $^{246}$Fm the width of the calculated distribution is significantly smaller than that of the experimental distribution. The latter indicates that a more sophisticated way to account for non-equilibrium effects is needed. This requires further theoretical developments, which are underway.

\label{sec:acknowledgments}
\section{Funding}

This work was supported by the Russian Foundation for Basic Research (project no. 18-52-15004), and the Joint Institute for Nuclear Research (grant no. 22-502-06). Т.М.S. was partly supported by the Kazan Federal University Strategic Academic Leadership Program.

\printbibliography

\end{document}